\documentclass[aps,prl,preprint,groupedaddress]{revtex4-1}

\usepackage{graphicx}
\usepackage{dcolumn}
\usepackage{bm}

\def\ia{\' \i}

\begin{document}

\bibliographystyle{prsty} 

\title{Low energy pathways for reproducible {\it in vivo} protein folding}

\author{L. Cruzeiro}

\thanks{Most of these simulations were performed at the Milipeia cluster of the Laboratory for Advanced Computing of the University of Coimbra, Portugal.}
\affiliation{CCMAR and FCT, Universidade of Algarve, Campus de
Gambelas, 8005-139 Faro, Portugal}

\date{\today}

\begin{abstract}

Two proteins, one belonging to the mainly $\alpha$ class and the other belonging to the $\alpha$/$\beta$ class, are selected to test a kinetic mechanism for protein folding. Targeted molecular dynamics is applied to generate folding pathways for those two proteins, starting from two well defined initial conformations: a fully extended and a $\alpha$-helical conformation. The results show that for both proteins the $\alpha$-helical initial conformation provides overall lower energy pathways to the native state. For the $\alpha$/$\beta$ protein, 30 \% (40\%) of the pathways from an initial $\alpha$-helix  (fully extended) structure lead to unentangled native folds, a success rate that can be increased to 85 \% by the introduction of a well-defined intermediate structure. These results open up a new direction in which to look for a solution to the protein folding problem, as detailed at the end.

\end{abstract}



\maketitle

\section{Introduction.}

Proteins are mega molecules. Even a small protein, with 60 amino acids, possesses approximately one thousand atoms which are linked by covalent bonds, hydrogen bonds, electrostatic interactions and van der Waals interactions. Despite the inherent complexity of such a system, proteins, in cells, are capable of folding reproducibly to a well defined three-dimensional structure that is generally irregular and lacks symmetry. How do proteins do it? According the thermodynamic hypothesis \cite{anf73} the native structure of proteins corresponds to the minimum of the free energy and protein folding is a thermodynamic process simply driven by thermal agitation. The complementary proposal that the free energy landscape is funnel-shaped was later put forward to explain also why folding can proceed comparatively fast \cite{bry,onu97,dc97}. However, in spite of many decades of study, and of the improvements in the accuracy of force fields \cite{gromos,amber,charmm}
and in the power of computer facilities \cite{blue,folding_at_home}, it has not yet been possible to determine a protein structure solely from its amino acid sequence and the thermodynamic and funnel hypotheses remain to be proven.

Here another possibility is considered, namely, that protein folding is a kinetic process and that the native structure is just one of the many kinetic traps in which proteins can find themselves in \cite{jbp01}. In previous works \cite{arxiv,jpoc08,molphys09,spr10} it has been shown that a protein from a given CATH \cite{cath} class can be forced into artificial, non-native structures that belong to other CATH classes and maintain these structures for at least 50 nanoseconds. These results favour instead a {\em multi-funnel-shaped} free energy landscape of proteins and a specific {\em kinetic} process for {\it in vivo} protein folding was suggested \cite{spr10}. For a kinetic process to lead to a well-defined three dimensional structure everytime, it must be associated with a specific pathway, as Levinthal first proposed \cite{lev68}; however, and equally importantly, {\em it must always start from the same well-defined initial conformation}. In \cite{spr10} it was suggested that this specific conformation, that is, the conformation that {\em all} proteins have immediately after synthesis, is helical and that the first step in folding is the bending of this initial helix at specific amino acid sites. The purpose of the present study is to make a preliminary test of the efficiency of such a kinetic process for folding. To that end, two proteins, representative of the mainly $\alpha$ and $\alpha/\beta$ CATH \cite{cath} protein classes, were selected and pathways from the initial conformation to the native state were generated using Targeted Molecular Dynamics (TMD) \cite{sek93}. In TMD simulations, harmonic restraints are added to the protein force field in order to drive an initial protein conformation to a given target conformation \cite{sek93,afc99}. TMD simulations were first applied to the T $\leftrightarrow$ R transition of the protein insulin \cite{sek93} and have since been used to study a variety of other problems, including, e.g., the conformational changes associated with the functioning of a molecular motor \cite{otk10}, the elucidation of the reaction steps in the full catalytic cycle of a protein involved in an electron transfer process \cite{trxr10}, as well as in protein folding \cite{fac00}. Thus, TMD simulations have generally shown their usefulness to model processes that involve large conformational changes.

Experimentally, it has not yet been possible to determine the structure of nascent chains, that is, of the structure proteins have as they come out of the ribosomal tunnel but it is known that both fully extended and $\alpha$-helical structures fit into the tunnel dimensions \cite{ecc09}. In the present study, TMD simulations are used to compare the efficiency of folding from an initial helical conformation, proposed previously to be the conformation of proteins immediately after synthesis \cite{spr10} to the efficiency of folding from a fully extended conformation.

Two proteins, each representative of one of the four main protein CATH classes \cite{cath}, were chosen and their native structures were obtained from the Protein Data Bank (PDB) \cite{pdb}. One of these proteins (PDB1BDD  \cite{1bdd}, 60 amino acids, 941 atoms, mainly $\alpha$) has a native structure constituted by three $\alpha$-helices, while the second protein (PDB1IGD \cite{1igd}, 61 amino acids, 927 atoms, $\alpha$/$\beta$) has a native structure that includes one $\alpha$-helix and a $\beta$-sheet. Two initial conformations, both of which have been identified as viable experimentally \cite{ecc09}, were taken for each of the two proteins, namely, one conformation in which the backbone is fully extended and a second conformation in which the backbone is folded into an ideal $\alpha$-helix. The energy minimized versions of these two conformations were used as initial conditions for the TMD simulations.
The ff99SB force field \cite{ff99SB}, with an implicit solvent, implemented in AMBER 9 \cite{amber9}, which has been shown to give a better representation of secondary structure \cite{ff99SB_09}, was used in all simulations. For each initial conformation, 20 independent TMD trajectories to the native state were generated by changing the seeds of the random forces in the Langevin thermal baths, with $T=298$~K in all cases. In TMD simulations the harmonic forces that drive the initial conformations to the native structure arise from the following term added to the atomic potential energy function:
\begin{equation}
U_{\rm TMD}= \frac{1}{2} \, k \, N \, \left( {\rm RMSD}_N (t) - {\rm RMSD_{\rm target}} (t) \right)^2
\label{Utmd}
\end{equation}
where $N$ is the number of atoms used to calculate the root mean square deviation per atom (RMSD), ${\rm RMSD}_N (t)$ being the RMSD, with respect to the native structure, of the conformation the protein has at time $t$, and ${\rm RMSD_{\rm target}} (t)$ being the target RMSD at that time. In the simulations reported here, $k=100$ kcal/mol/{\AA}$^2$ and $N$ is the total number of carbons, nitrogens and oxygens in the backbone of the two proteins selected.

In TMD simulations, ${\rm RMSD_{\rm target}} (t)$ is a linear function of time, being completely defined by the values at two time instants. Here, these two times are 1) its initial value,  ${\rm RMSD_{\rm target}} (t=0)$, given in table \ref{one}, and 2) its final value, set to $0.1$ {\AA} in all trajectories. As shown in table \ref{one}, for both proteins, the $\alpha$-helical conformation is more than $2.4$ times closer to the native structure than the fully extended conformation.
\begin{table}
\caption{${\rm RMSD_{\rm target}} (t=0)$, in {\AA}, for the two initial conformations of the two proteins. \label{one}}
\begin{ruledtabular}
\begin{tabular}{ccc}
PROTEIN & INITIALLY EXTENDED & INITIALLY $\alpha$-helix \\
mainly $\alpha$ (1BDD) & 55.78 & 21.54 \\
$\alpha$/$\beta$ (1IGD) & 59.76 & 24.46 \\
\end{tabular}
\end{ruledtabular}
\end{table}
Preliminary simulations starting from the $\alpha$-helical conformation showed that the total potential energy (including the TMD term) does not vary much until ${\rm RMSD}_N \approx 6$ {\AA} and so, in order to make the simulations with the two different initial conformations as equivalent as possible, in all the TMD simulations presented here, two  ${\rm RMSD_{\rm target}} (t)$ functions were used, i.e., the first 0.1 ns were spent in the convergence of the initial conformation to within 6 {\AA} of the native structure, that is, ${\rm RMSD_{\rm target}} (t=0.1 \; {\rm ns})  = 6$  {\AA}, and 0.4 ns were allowed for the further convergence to within 0.1 {\AA} of the native conformation. In this way, a slower rate of change of the RMSD is imposed in the final stages of convergence to the native state, allowing more time for the side chains to avoid steric overlaps, and, most importantly, making the rate of RMSD change in this latter stage the same for all simulations.
%
The overall duration of each TMD simulation was 0.5 ns, comparable to other TMD protein folding simulations \cite{fac00}.

\begin{figure}
 \begin{center}
    \includegraphics[width=0.8\textwidth]{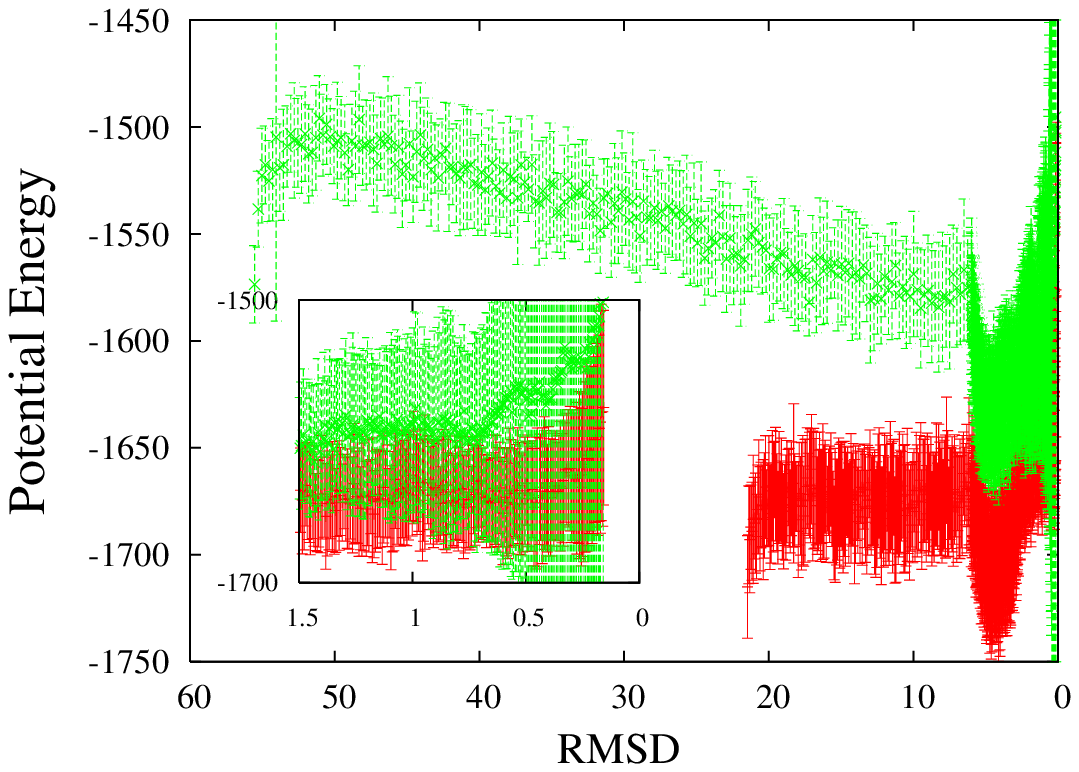}
  \end{center}
  \caption{}
  \label{ep_1bdd}
\end{figure}

Each of the two initial conformations was taken as an initial condition in 20 independent TMD simulations, as detailed above. Figure \ref{ep_1bdd} shows the average over the 20 trajectories and the corresponding standard deviations of the total potential energy as a function of instantaneous value of ${\rm RMSD}_N $ for the mainly $\alpha$ 1BDD protein \cite{1bdd}. The curve in red is for the trajectories starting with a $\alpha$-helical conformation and the curve in green is for the trajectories starting with a fully extended backbone conformation. Included in the values displayed in figure \ref{ep_1bdd} is the contribution of the TMD term (\ref{Utmd}) which, however,
%
 only starts to rise above 0.3 \% of the total value when the RMSD distance to the native structure becomes less than 0.6 {\AA} (not shown). This means that the energetics of these folding pathways is determined essentially by the atomic interactions in the ff99SB AMBER potential \cite{ff99SB}.

Figure \ref{ep_1bdd} shows that the pathways from the initially extended conformations are populated by protein conformations which have, on average, a potential energy that is much greater than those that arise when the initial conformation is $\alpha$-helical; indeed, even in the final stages of approach to the native state, the former protein conformations are at least 40 kcal/mol above the latter. Furthermore, contrary to what was found in previous folding simulations \cite{fac00} and below for the $\alpha$/$\beta$ 1IGD protein, all pathways of the mainly $\alpha$ 1BDD protein lead to conformations with a perfectly folded backbone at least within 0.3 {\AA} of the native structure.

%
%

\begin{figure}
 \begin{center}
    \includegraphics[width=0.6\textwidth]{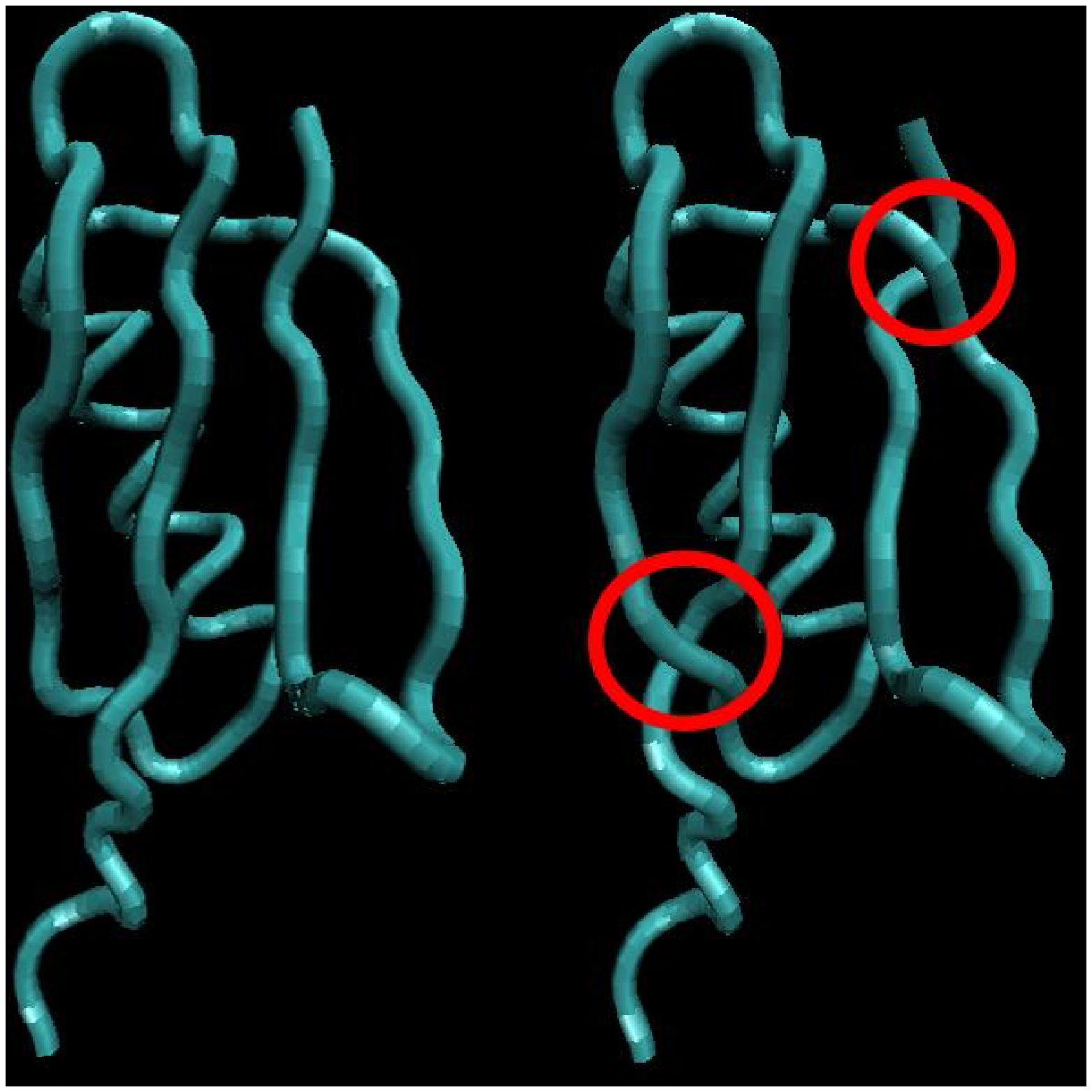}
  \end{center}
  \caption{}
  \label{1igd_ent}
\end{figure}

\begin{figure}
 \begin{center}
    \includegraphics[width=0.8\textwidth]{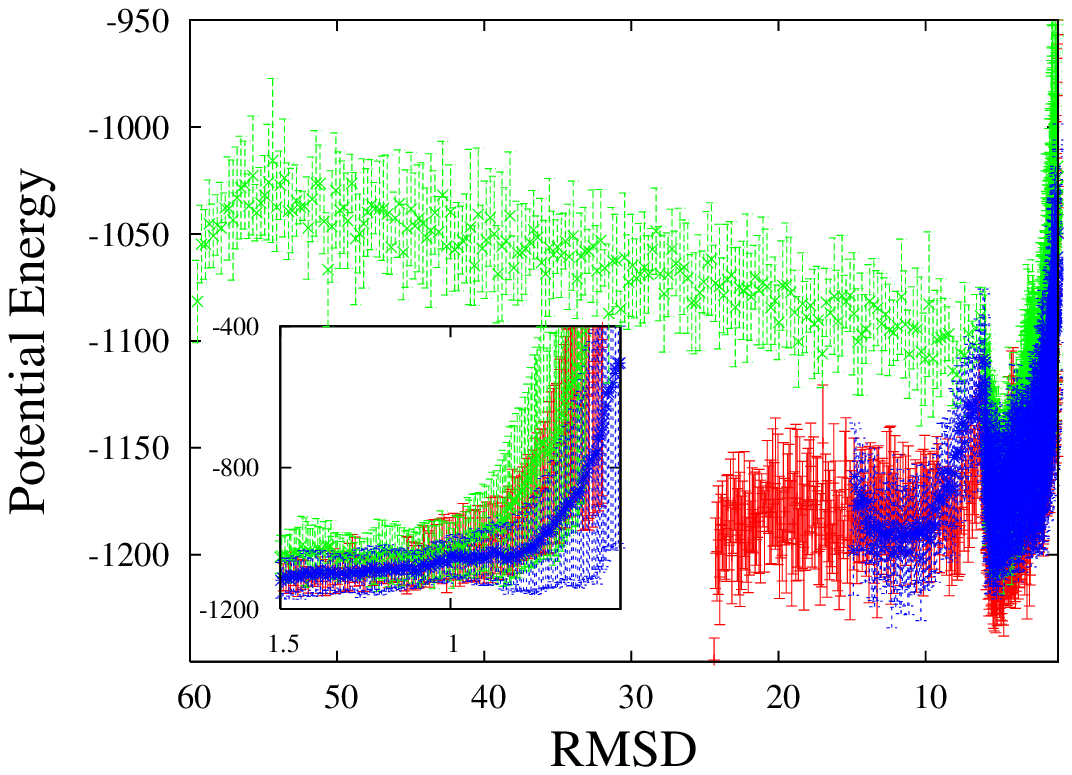}
  \end{center}
  \caption{}
  \label{ep_1igd}
\end{figure}

To test further the hypotheses put forward in \cite{spr10}, another protein, the $\alpha$/$\beta$ 1IGD protein \cite{1igd}, whose native structure includes both an $\alpha$-helix and a four stranded $\beta$-sheet, was used. As reported previously \cite{fac00}, many of the TMD folding simulations of the $\alpha$/$\beta$ 1IGD protein lead to final structures that, although apparently close to the native structure, differ from it by entangled backbone folds such as those highlighted by the red circles in figure \ref{1igd_ent}. Indeed, inspection of the trajectories with the Visual Molecular Dynamics (VMD) software \cite{vmd} reveals that 12 (14) of the 20 TMD simulations starting with a fully extended ($\alpha$-helical) conformation lead to such entangled final structures.
%
%
%
These entanglements can be resolved by letting the chains go through each other, an artificial process enabled by the TMD term (\ref{Utmd}) but of course penalized by extremely large values of the potential energy. 
Thus, in figure \ref{ep_1igd} the averages are made over viable pathways only, that is, over the TMD trajectories in which such backbone entanglements did not occur. 
Although the TMD term contributes a little more than before to the energetics of the pathways, its total amount only rises above 1 \% when the RMSD distance to the native structure goes below 1 {\AA}, with the larger values being associated with entangled structures (not shown). Figure \ref{ep_1igd} shows that, also for the $\alpha$/$\beta$ 1IGD protein, the viable pathways from an initial $\alpha$-helix to the native state are much less energetic than the corresponding ones from a fully extended structure, with the conformations covered by the latter having a potential energy more than 130 kcal/mol greater than the former to start with, and with their values only merging when the RMSD distance to the native structure is approximately 6 {\AA}.

%
%

\begin{figure}
 \begin{center}
    \includegraphics[width=0.6\textwidth]{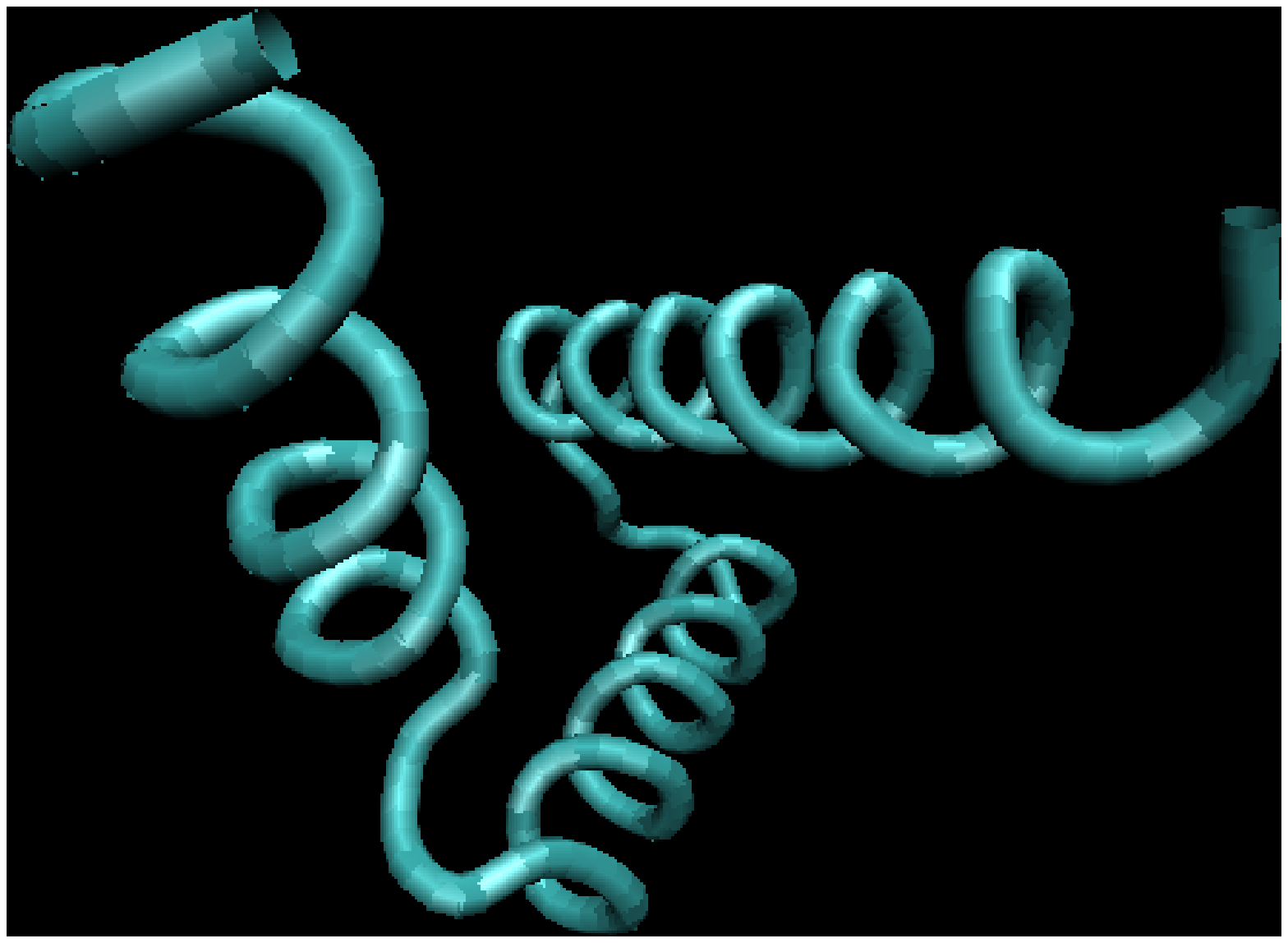}
  \end{center}
  \caption{}
  \label{embrio}
\end{figure}

In \cite{spr10} it is suggested that the initial conformation of a protein is a helix and that the first step in protein folding is the bending of this helix at specific amino acid sites. In the case of proteins whose secondary structure is just a set of helix bundles, as in the mainly $\alpha$ 1BDD protein \cite{1bdd}, this first step is the only major step in their folding pathways. VMD \cite{vmd} animations of the TMD trajectories of the 1BDD protein do show pathways of this sort when the initial conformation is $\alpha$-helical (not shown). (When the initial conformation is fully extended, on the other hand, the formation of the $\alpha$-helices 
constitutes the last step, coming after the backbone has folded to the native topology). For proteins whose native structure includes both helices and sheets, as is the case of the $\alpha$/$\beta$ 1IGD protein, it was further suggested \cite{spr10} that the formation of the $\beta$-sheets is due to destabilizing interactions between the amino acid side chains that are thus thrown together by the first step. VMD \cite{vmd} animations of the TMD trajectories of the $\alpha$/$\beta$ 1IGD protein starting from the $\alpha$-helical conformation reveal pathways that start off in that manner but in which the helical portions very quickly become distorted because the convergence to the right secondary structure is mixed with the convergence to the right backbone fold (not shown). This is to be expected because of the unspecific character of the TMD term (\ref{Utmd}). In order to test, in a more direct way, the effect that an intermediate all-helical conformation has on the folding efficiency of the $\alpha$/$\beta$ 1IGD protein, such a putative intermediate was generated by substituting the $\beta$-sheets in the native structure of I1IGD by $\alpha$-helices, as shown in figure \ref{embrio}. This intermediate shall here be called an {\em embrio} because, according to the kinetic process proposed in \cite{spr10}, its structure is one of the main determinants of the final, native, structure of the protein.

%
%

Further TMD simulations were run by first driving an initial $\alpha$-helix to the embrio, and then by driving the embrionic conformations
to the native structure of 1IGD. As the RMSD between the $\alpha$-helix and the embrio is 23.78 {\AA} and the RMSD between the embrio and the native 1IGD structure is 15.05 {\AA}, in order to keep approximately the same rate of change of RMSD as before, the first simulations had a duration of 0.05 ns and the second simulations had a duration of 0.45 ns. 
In the first 0.05 ns all 20 trajectories converge to within $0.1$ {\AA} of the embrio structure; furthermore,
17 out of the 20 trajectories from the embrio to the native state lead to un-entangled final structures (not shown). Figure \ref{ep_1igd}, in which the potential energies of the protein conformations in these 17 viable folding pathways are displayed in blue, shows that the $\alpha$-helix $\rightarrow$ embrio $\rightarrow$ native trajectories provide, on average, the lowest energy folding pathways for the $\alpha$/$\beta$ 1IGD protein.


This study does not prove the hypothesis that the $\alpha$-helix is the conformation that all proteins have as they come out of the ribosomal tunnel; it does not prove either the hypothesis that the first step in the folding of all proteins is the formation of a compact core, here called an embrio, constituted only by helices (and disordered helices or turns), nor does it prove that $\beta$-sheets arise when one or more of the helices in that embrio is not stable; but it does show that those hypotheses are viable in that they can lead to low energy folding pathways to correctly folded, unentangled, native states. The simulations reported here also show that, of the two nascent chains that have been identified as experimentally feasible \cite{ecc09}, that in the form of a helix is to be preferred to a fully extended conformation. Furthermore, the results here provide a glimpse of the many entangled states that can be expected to arise if protein folding is just a random search of the native structure, driven by thermal noise, especially if the initial structure is also arbitrary. Indeed, only 30 \% and  40 \% of the direct pathways to the native state, from a $\alpha$-helical and a fully extended initial conformation, respectively, lead to a correctly folded 1IGD protein, even in the presence of the artificial TMD terms (\ref{Utmd}). On the other hand, the simulations here also show how a well defined intermediate/transient structure in the folding pathway can dramatically increase the odds of reaching the native structure in a reproducible manner: with the roughly built embrio  tentatively considered here (see figure \ref{embrio}) the probability of reaching the native structure became 85 \%!
{\em The really important point, though, is the new direction in which to look for a solution to the protein folding problem that is opened up by these and other \cite{arxiv,jpoc08,molphys09,spr10} results: instead of a thermodynamic process, a kinetic process; instead of arbitrary initial structures, 
a well defined and already ordered nascent chain; instead of many pathways, or even preferred pathways, a specific pathway for folding.} Another important point is the concrete program of research that naturally follows from it \cite{spr10}: 1) the identification of the key amino acids, or amino acid sequences, at which the bends of the initial $\alpha$-helix occur, can be made by a statistical analysis of the known structures of proteins formed by helical bundles only; 2) 
the same structures can be mined to find the relative orientations of the helical pieces in their respective embrios, and 3) once the key amino acids and these initial orientations have been identified, a statistical analysis of the known structures of $\alpha$/$\beta$ proteins may tell us which amino acid sequences in these initial helical pieces generate repulsive side chain interactions that, in turn, destabilize those helices and lead to the formation of $\beta$-sheets. It is indeed hoped that the results here will inspire further work to explore this new direction.


\newpage

\begin{center}
CAPTION LIST
\end{center}

\begin{enumerate}
\item (color online) Average total potential energy (kcal/mol) over 20 independent TMD simulations, together with the corresponding standard deviations, as a function of the RMSD distance to the native structure of the mainly $\alpha$ 1BDD protein \cite{1bdd}. The starting conformation is fully extended ($\alpha$-helical) for the green (red) curves. The inset shows the final stages of convergence to the native structure.


\item (color online) The native fold of the $\alpha$/$\beta$ 1IGD protein is depicted on the left and on the right the red circles highlight two of the most common backbone entanglements that arose in the TMD simulations of this protein. The figures were made with the VMD software \cite{vmd}.

\item (color online) Average total potential energy (kcal/mol) over non-entangled TMD simulations, together with the corresponding standard deviations, as a function of the RMSD distance to the native structure of the $\alpha$/$\beta$ 1IGD protein \cite{1igd}. The starting conformation is fully extended ($\alpha$-helical) for the green (red) curves and the embrio (see text and figure \ref{embrio}) for the blue curves. The inset shows the final stages of convergence to the native structure.


\item (color online) The putative embrio (see text) built from the native fold of the $\alpha$/$\beta$ 1IGD protein by substituting the native $\beta$-sheets by $\alpha$-helices. This figure was made with the software VMD \cite{vmd}.

\end{enumerate}

\newpage

\newpage

\newpage

\newpage

\end{document}